\documentstyle {article}
\input epsf
\begin{document}
\vspace*{-1cm}
\begin{flushright}

                                                                                                KEK CP - 044\\ 
                        KEK Preprint 96 - 16    \\
CPTH S448.0496  \\
\end{flushright}
\renewcommand{\baselinestretch} {1.5}
\large

\begin{center}
{\Large {\bf

Equivalence Between Vector Meson Dominance and Unitarised
 Chiral Perturbation Theory }}
\vskip 0.3in

Le viet Dung$^{({b,c})}$ and Tran N. Truong$ ^{({a,b})}$
\vskip 0.2in
$^{(a)}${\it National Laboratory for High Energy Physics (KEK)
\vskip 0.03in
Oho 1-1 Tsukuba, Ibaraki 305, Japan
\vskip 0.2in
$^{(b)}$Centre de Physique Theorique de l'Ecole Polytechnique
\vskip 0.03in
91128 Palaiseau, France
\vskip 0.2in
$^{(c)}$ Institute of Physics
\vskip 0.03in

Hanoi, Viet Nam}

{\bf( Revised Version )}

\end{center}
\vskip 0.3in
\begin{center}
{\large\bf Abstract}
\end{center}

        It is  explicitly shown that either the approximate solution of the 
integral equation for the inverse of the pion form facto,r or the result of the Pad{\'e}
 approximant method of resumming the one loop Chiral Perturbation Theory
 (CPTH) are equivalent to the standard vector meson dominance (VMD) models,
 using the vector meson coupling to two pseudoscalars given by the KSRF 
relation. Inconsistencies between the one loop CPTH and its unitarised version
  (or the VMD model) are pointed out. The situation 
is better for the CPTH calculation of the scalar form factor and the related
 S-wave $\pi \pi$ scattering. The branching ratios of $\tau \to \pi^+ \pi^0 \nu $,
  $\tau \to K \pi \nu $,
   $\tau \to K^+ \eta \nu$ and $\tau \to K^+ \bar{K^0} \nu$
 using only two inputs as the $\rho $ and $K^*$ masses, or the two corresponding 
 rms radii,
 agree  with the experimental data. Using the same number of parameters,
 the corresponding one loop CPTH calculation
 cannot explain the $\tau$ data.

\eject

\noindent

\section{Introduction}

        The hypothesis of Vector Meson Dominance (VMD) \cite{Nambu} has proved to be an
 useful and convenient concept in low energy hadronic physics.
 It enables us to describe many low energy phenomena, below 1 GeV scale,
 in a compact and convenient language, although not always correct.
 An apparently different method was
 based on the dispersion relation approach which was the main activities 
of the soft hadronic physics in the fifties. It was soon realised to some
authors that
 the VMD model can conveniently be used to describe the more complicated 
dispersive approach.

        A more recent approach to these problems with a different goal 
was based on the Chiral Perturbation Theory (CPTH) pioneered by Li and Pagels \cite{Li},
 Lehmann \cite{Lehmann}, Weinberg \cite{Weinberg} and Gasser and Leutwyler
 \cite{GL1,GL2,Don}. In the Lehmann's
 approach, the unitarity relation of the S-matrix was taken into account but 
not the chiral symmetry breaking effect. In the 
   Weinberg approach, the question of unitarity was ignored
 giving place to the more systematic approach of the perturbation theory in which
the unitarity relation was only satisfied perturbatively, i.e. order by order, but 
the chiral symmetry breaking effect was taken into account. The latter approach 
enables us to derive systematic low energy theorems, provided that strong
 interaction can be treated perturbatively, which is an assumption
 and has to be demonstrated.

In the late fifties, the
 problem of the validity of the perturbation theory for strong processes
 was questioned, but this issue has been ignored in recent numerous studies
of Chiral Symmetry using CPTH, either by {\em assuming} that the strong
 interaction involved was not sufficiently strong to invalidate 
the perturbative approach, or that the CPTH was effectively a low energy power series
 expansion in momentum which could evade the unitarity constraint.

 There are  few  publications on Chiral Symmetry where 
 chiral symmetry breaking effect
 and unitarity are simultaneously taken 
into account. This last approach, combining the CPTH with Unitarity, will
be called as the Unitarised Chiral Perturbation Theory (UCPTH). It consists
 in deriving a similar expression as the one loop CPTH result but is supplemented
by  using 
 the inverse amplitude,    
the $N/D$  or the Pad{\'e} approximant methods in order to satisfy the 
elastic unitarity relation \cite{Lehmann,Truong1,Truong2,Truong3,Jhung}. 
This procedure enables us
 to extend the perturbation theory to incorporate low energy resonance or
 bound state phenomena which the standard CPTH cannot handle. We can, in principle,
 include the inelastic effects but calculation becomes less simple. Except for a most
 simple calculation of the inelastic effect, we deal in this article the low energy phenomena
 where the elastic unitarity relation dominates the calculation.

The elastic unitarity relation for the form factor,
 which is incorporated in the UCPTH approach,
 leads to the following two consequences. First, the expression for the form factor is 
an integral equation of the Muskhelishvilli-Omn{\`e}s type \cite{Omnes} \cite{Pham2} 
which should not be solved perturbatively. Second, the phase of the form factor 
 in
the low energy region (where the inelastic effect can be neglected), must
be the same as that given by the corresponding strong interaction phase shift \cite{Watson}.
 The solution of the 
Muskhelishvilli-Omn{\`e}s integral equation \cite{Omnes} has a polynomial ambiguity,
 to be discussed below, but the phase theorem is independent of  such an ambiguity.
 Comparision between theory and experiment should therefore be done with the magnitude
{\em and } phase 
of the form factor.

The one loop CPTH calculation for the vector and scalar pion form factor
 was done a long time ago \cite{GL1}.
 A 2 loop CPTH calculation was also recently done \cite{Gasser3,Colangelo}. These papers treated 
the form factors perturbatively and therefore cannot be used to calculate the 
$\tau$ decay to 2 pseudoscalars which is dominated by the $\rho$ and $K^*$ resonances. 
 
        In a  recent paper on the study of the isovector form factors\cite{Truong1},
 it was shown that the question of the unitarity must be respected in order
 to describe in a simple manner the low
 energy pion physics with or without resonances. It was explicitly pointed
 out that two approaches could be used: a) One could use the standard CPTH
 but after having done the one loop perturbative calculation, one must resum the series by the 
Pad{\'e} method in order to satisfy the elastic unitarity relation  
 \cite{Lehmann,Truong1,Truong2,Truong3,Jhung} (this approach could be regarded as the large 
number of flavor $N_f$ expansion).  
 Other applications of the Pad{\'e} method in physics have been shown to
 be successful \cite{Samuel}, but we give here the reason why it should be.  b)
 One could write the dispersion relation of the inverse of the form 
factors or the scattering amplitudes which takes automatically
 into account of the elastic unitarity relation, and then solve approximately the resulting integral
 equation \cite{Truong1,Hannah}.
 The well-known $N/D$ method could be regarded as an extension
 of this method but is more flexible.

         Both methods yield the same result and give rise to the well-known
 KSRF relation for the width of the vector meson \cite{KSFR} . It was implicitly
 shown that these two methods are equivalent to the VMD model \cite{Truong1,Beldjoudi1,Beldjoudi2}  using
 the $\rho\pi\pi$ coupling as given by the KSRF relation \cite{KSFR}. The 
present method contains less parameter than the VMD model because the KSRF 
relation is a direct consequence of our approach and not an added assumption.

        In this paper we want to point out the equivalence between the
 VMD model on one hand, and the Pad{\'e}, the inverse amplitude and the N/D
 methods on the other hand, for the calculation of
 $\pi\pi$, $K \bar{K},\pi K$ and $\eta K$ vector form factors.
 They have
 the same number of parameters as those used in the one loop CPTH method but
 yield a much more significantly improved prediction due to 
the respect of the constraint of the unitarity (by a factor of more than 50 
 in the $\pi K$ form factors squared) \cite{Truong1,Beldjoudi1,Beldjoudi2}.

 We  want to improve the inverse amplitude method, by expressing the
 form factor in terms of the N-function of the N/D method for the corresponding
 strong amplitude. We briefly give the reason why the N function used in the
 vector pion form factor calculation
 can be approximated by the Weinberg low energy expression.

 As mentioned above, the unitarity relation for the form factor would require that  its
 phase must be that of the corresponding strong interaction, if the
 inelastic effect was unimportant in the physical region of interest \cite{Watson}.
This is the only model independent relation which we can make, apart
 from the (more fundamental and rigourous)
 Ward identity on the  the non renormalisation of the charge of the pion at 
 the zero momentum transfer. For the vector and scalar pion form factors,
because there are no significant inelastic effect below 0.9 $GeV^2$, the phase theorem
should be valid in this region. Hence the phase of the vector pion form
 factor should have a phase
 of $90^0$ at 770 MeV, the  $\rho$ vector meson mass, and similarly, the $\pi$K vector
 form factor has a phase of $90^0$ at the $ K^*$ mass. Throughout this article 
and in the UCPTH approach, {\em we have to rely on this phase theorem
to check the validity of our approximation scheme}. This point is ignored  in the
 standard CPTH calculations.

 As a direct application of the UCPTH method, we show that the branching ratios of
 $\tau \rightarrow  Vector Mesons +\nu$  agree reasonably well with the experimental data
 using only inputs as the $\rho $ and $K^*$ masses, or 
alternatively, the corresponding rms radii 
 ( which the usual CPTH cannot be used or is wrong, being too low, by a factor of 10
 to 50).
 The branching ratios of
  $\tau \to K^+ \eta \nu$ and $\tau \to K^+ \bar{K^0} \nu$ are however smaller than
 the observed branching ratio  by a factor of 2. A much better agreement
 with the data is obtained by introducing the polynomial 
ambiguity as a simulation of the inelastic effect, e.g. the contribution of 
the $\omega \pi$ intermediate state 
 in the unitarity relation. By fitting
 the height of the pion form factor data at the $\rho$ resonance 
 with the first order polynomial, 
the r.m.s. radius is now in good agreement with the data.
 Alternatively, using the experimental r.m.s. radius as an input, 
the magnitude of the pion form factor
 at the $\rho$ resonance is in a very good agreement with the experimental data.

 We also want to show that the inelastic effect of the $K \bar{K}$ states
 on the pion form factor, the inelastic effect of the $\eta K$ intermediate
 state on the $\pi K$ form factors are however negligible. A more correct approach, using the 
coupled two-channel unitarity calculation, will be done in the future.

     From a more fundamental viewpoint, we question the assumption made in the CPTH approach
 that perturbation theory can be used to study the vector pion form factor which could involve
  important non-perturbative effects. For 
this purpose, we want to point out  some novel
 theoretical inconsistencies
 between the one loop CPTH and its unitarised version UCPTH calculation 
 of the vector pion
 form factor (or equivalently the VMD model); the same
 observation is also valid for the
 P-wave $\pi \pi$ elastic scattering. Our observation on the inconsistency can be extended to
 the two-loop CPTH calculation, and is conjectured to  higher orders.
 We point out, however,
that the problem of unitarity may not be serious
 in the CPTH calculation of the low energy
 I=0 scalar form factor or the related elastic
S-wave $\pi\pi$ scattering.

 It is our feeling that, not only because 
the Breit Wigner form cannot be expanded in a convergent power series of momenta  
near the $\rho$ resonance, because of this inconsistency,
 even with higher order loop calculation, CPTH will
remain as a incomplete low energy perturbative theory and cannot handle resonance or bound state 
problems which are a manifestation of the non perturbative effects.

The plan of this paper is organised as follows: In section 2, which is the main part of our 
paper, we give a
 detailed calculation of the vector pion form factor by perturbation method and also by 
the non perturbative inverse amplitude and the N/D methods where the elastic unitarity is
satisfied. In section 3 the scalar form factor
 is calculated also by two methods. We then point out that the use of the perturbation 
method can be justified here. In section 4, the equivalence
 between the UCPTH and the VMD is shown. Section 5 deals with the SU(3) generalisation of
 the form factors. Section 6 is devoted to the applications of the above calculations to the 
$\tau$ decays.

\section{Calculation of the Vector Pion Form Factor}

 {\bf a) Theoretical Consideration: One Loop CPTH Result  }

        We begin first by recalling  the main features of the vector pion form factor
 calculation using CPTH \cite{GL1,GL2}. Let us define the vector pion form factor as:

\begin{equation}
<\pi^+(p_1)\pi^-(p_2)\mid V_\mu^3(0)\mid0>= i(p_2-p_1)_\mu V (s)
\end{equation}

 where s is the
 momentum transfer squared $s=(p_1+p_2)^2$ and $V(0) = 1$. Using the dimensional
regularisation scheme, the one loop CPTH result is \cite{GL1}:

\begin{equation}
                 V^{pert.}(s) = 1 + {2\alpha_9^r(\mu)\over f_\pi^2} s + {1\over 96\pi^2f_\pi^2} ((s-4m_\pi^2)
H_{\pi\pi}({s}) - sLog{m_\pi^2 \over\mu^2} - {s\over 3} ) \label{eq:pert}
\end{equation}
 
where s is the the momentum transfer, $\alpha_9^r(\mu)$ is one of the renormalized
 constants defined by Gasser and Leutwyler \cite{GL1}, $f_\pi$=93 MeV, the experimental
 value of the pion decay constant, and

\begin{equation}
   H_{\pi\pi}(s) = (2 - 2 \sqrt{s-4m_\pi^2\over s}Log{{\sqrt{s}+\sqrt{s-4m_\pi^2}\over
2m_\pi}})+i\pi\sqrt{s-4m_\pi^2\over s} \label{eq:H}
\end{equation}

for $ s>4m_\pi^2$; for other values of s, $H_{\pi\pi} (s)$ can be obtained by analytic
 continuation. 

 We can explicitly
 introduce the expression for
r.m.s radius of the vector form factor in Eq. (\ref{eq:pert}).
 Using the definition $V ^{'}(0) =\frac{1}{6} r_V^2 = 1/s_{R_1}$ we have:

\begin{equation}
\frac{1}{s_{R_1}}=\frac{2\alpha_9^r(\mu)}{f_\pi^2}-\frac{1}{ 96\pi^2f_\pi^2}(
Log\frac{m_\pi^2}{\mu^2}+1) \label{eq:rms}
\end{equation}

Using this expression in Eq. (\ref{eq:pert}), we have:

\begin{equation}
                 V^{pert.}(s) = 1 +\frac{s}{s_{R_1}}\ + {1\over 96\pi^2f_\pi^2} ((s-4m_\pi^2)
 H_{\pi\pi}({s}) + {2s\over 3} ) \label{eq:pertv}
\end{equation}

It should be noticed that the real part of the sum of the last two terms on the R.H.S 
of Eq. (\ref{eq:pertv}) behaves at low energy as $s^2$. In using the r.m.s. radius as
 an experimental input to renormalise the calculation of $V^{pert.}(s)$,
 we can calculate perturbatively the vector form factor to the order of $s^2$ (modulo
 of a logarithm). In the language of the dispersion relation approach, the r.m.s
 radius of the pion vector form factor is used as the subtraction
constant.

\vskip 0.1 in
{\bf b) Non Perturbative Approach: Elastic Unitarity Constraint }

At issue here is whether the perturbative hypothesis used in the CPTH approach justified. We 
shall discuss this problem later. At the moment, we shall
 discuss the non perturbative problem by using the integral equation approach. Let us review
some fundamental properties of the form factor V(s). Because of the Ward identity and
 the analytic property of the form factor, 
we can write the following disperssion relation for V(s):

\begin{equation}
V(s) = 1+\frac{s}{s_{R_1}} +\frac{s^2}{\pi}\int_{4m_\pi^2}^\infty{
\frac{\sigma(z)}{z^2
(z-s-i\epsilon)}dz } \label{eq:ff}
\end{equation}

where an extra subtraction is made for convenience, and the spectral function
 $\sigma(z)$ must be taken to be real from time reversal invariance.

        The Muskhelishvilli-Omn{\`e}s integral equation \cite{Omnes}
 is obtained by  using the elastic unitarity  condition for the spectral function: 

\begin{equation}
V(s) = 1+\frac{s}{s_{R_1}} +\frac{s^2}{\pi}\int_{4m_\pi^2}^\infty{
\frac{V(z)  e^{-i\delta_1(z)}sin\delta_1(z)}{z^2
(z-s-i\epsilon)}dz } \label{eq:MO}
\end{equation}

where $\delta_1$ is the P wave $\pi\pi$ phase shift and we have made 2 subtractions
 in the dispersion relation for later purpose. Because of the  of the reality
condition of the
spectral function, in the
 region where the elastic unitarity relation was valid, the form factor
 would have the same phase as that of the strong amplitude.
 This is the content of the Watson phase theorem \cite{Watson}.

 The solution for the integral equation (\ref{eq:MO}) is well-known \cite{Omnes}
 and can also be obtained perturbatively by an {\em infinite} iteration
 of the integral equation:

\begin{equation}
V(s) = P_n(s) exp(\frac{s}{\pi}\int_{4m_\pi^2}^{\infty}
{\frac{\delta_1(z)}{z
(z-s-i\epsilon)}dz)}\label{eq:Omnes}
\end{equation}

where $P_n(s)$ is the polynomial ambiguity with real coefficients,
 normalising to unity at s=0;
they could represent, at low energy, 
 uncalculable higher energy inelastic effects. We shall assume at the moment
 that $P_n(s)=1$
in the physical region of interest. As can be seen, the phase theorem
due to the unitarity of the S-matrix is automatically satisfied. An extension 
of this integral equation to include a phenomenological inelastic spectral function 
was previously considered  and its solution is also known \cite{Pham2} (see below).

The solution of the integral equation could have been
 guessed from the reality condition of 
spectral function: As mentioned above, from this condition, the phase of the form factor 
 is the phase of the strong amplitude.
 Because $e^{i\delta}$ is not an analytic function,
the only correct solution for $V(s)$ is given by Eq. (\ref{eq:Omnes}).
        The one loop CPTH solution, Eq. (\ref{eq:pertv}), is the {\em once} iterated solution of the 
integral equation Eq. (\ref{eq:MO}): This perturbative result can be obtained
 by setting V(s) in the integral 
of Eq. (\ref{eq:Omnes}) to be unity and $f_1^{tree}(s) = 
 e^{i\delta_1(s)}sin\delta_1(s)/\rho(s) =(s-4m_\pi^2)/(96\pi f_\pi^2)$
 which is purely real; 
this result follows from the evaluation of the one loop Feynman graph using the 
Cutkosky rule for the absorptive part and dispersion relation. The once iterated solution of the
Muskhelishvilli-Omn{\`e}s integral equation makes no sense, because the exact solution
 of this integral equation, as tated above, requires an infinite iteration.
This problem was discussed in reference \cite{Truong1}.

In order to compute the pion form factor, we can either use the experimental P-wave
 phase shift $\delta_1$ or the calculated $\delta_1$ phase shifts (obtained from the
construction of a unitarised P-wave amplitude) in Eq. (\ref{eq:Omnes}). To see the effect of the exact solution of the integral equation, let 
us observe that the P-wave $\pi\pi$ phase shifts passing through $90^0$ at the
$\rho$ mass, hence we approximate the P-wave phase shifts $\delta$  by
 $\pi\theta (s-s_\rho)$,
 the zero width approximation for the $\rho$ resonance; the exact solution
 obtained from Eq.(\ref{eq:Omnes}) is therefore $s_\rho /(s_\rho - s)$.
 It is clear from this example that we should
 develop a power series expansion for the inverse of the form factor,
 if we wanted to incorporate the low energy property of CPTH and at the same time a
 non perturbative result to take into account of a
possible resonant behavior. This was exactly
 done in reference \cite{Truong1} because the {\em inverse} amplitude has a nice property 
that the elastic unitarity relation for the $\pi\pi\rightarrow\pi\pi$ partial
 wave is automatically satisfied.

 The non-relativistic limit of this
 expansion was done a long time ago 
and is known as the
"effective range theory" \cite{Bethe}; it is a momentum power series
expansion of $kcot\delta$
which preserves automatically the unitarity of the S-matrix. This is so 
because the elastic unitarity of the partial wave  enables us to write
 $f(k)=e^{i\delta}sin\delta/k=1/k(1-icot\delta)$ , where k is the c.o.m. momentum,
hence the power series expansion in $kcot\delta$ which is the effective range expansion, 
is an expansion of the 
inverse of the amplitude. This type of expansion enables
 us to handle very well the low energy bound state (triplet S-wave) and the
 resonance (singlet S-wave)
 of the low energy nucleon-nucleon scattering.
 The missing part of the Hilbert transform or 
 analyticity  is presumably unimportant for a non-relativistic theory. 
( Unfortunately, standard explanations
of the low energy effective range theory do not emphasize the question of
  unitarity of the S-matrix which is much more important than the potential
  shape dependence of the expansion).
  
 As applied to the form factor problem, to the extent
 that the discontinuity of the left hand cut for the partial wave amplitude
 can be neglected,  or can be treated perturbatively  (see below),
 the pion form factor can straightforwardly be obtained by considering the
 integral equation for the inverse form factor amplitude \cite{Truong1}.
 This result is equivalent to resum the
 perturbation series using the infinite bubble summation of the pion loops.

Let us now carry out our analysis, 
without making any assumption on the left hand cut structure.
 Because the form factor has a cut from $4m_\pi^2$ to
infinity, so does its inverse, apart from the contribution coming from the zeros
of the form factor appearing as poles which we assume to be absent here. We shall later 
give a phenomenological description of these zeros as uncalculable
 inelastic effect which has not been taken into account here.

 The inverse of the form factor
$V^{-1}(s)$ satisfies also the same analytic property as the form factor $V(s)$, hence we
 can write a dispersion relation for it.
 Assuming that V(s) does not have a zero or its position is 
far from the physical region of interest, using the elastic unitarity condition,
we have:

\begin{equation} 
V^{-1}(s)=1-\frac{s}{s_{R_1}}\ -\frac{s^2}{\pi}\int_{4m_\pi^2}^\infty{
\frac{V^{*-1}(z)  e^{-i\delta_1(z)}sin\delta_1(z)}{z^2
(z-s-i\epsilon)}dz } \label{eq:inverse}
\end{equation}

From the general property of the analytic function, the partial wave amplitude
$f_1(s) =e^{i\delta_1}sin\delta_1/\rho(s)$ where $\rho(s)=\sqrt{1-4m_\pi^2/s}$,
$f_1(s)$ can always be written as the product of the 2 cuts, the right hand cut or
 the unitarity cut and the left hand cut due to the exchanged contributions:
 $f_1(s)=N(s)/D(s)$.
 Because we use the hypothesis of the 
elastic unitarity, the right hand cut function $D^{-1}(s)$ has the same phase
 representation as 
given by Eq.(\ref{eq:Omnes}) with the normalisation
 $D^{-1}(0) =1$ and where we set $P_n(s)=1$.
Hence we can set $V^{*-1}(s) e^{-i\delta(s)}sin\delta(s)=\rho(s)N(s)$.
Eq. (\ref{eq:inverse}) can now be rewritten as:

\begin{equation}
V^{-1}(s)=1-\frac{s}{s_{R_1}}\ -\frac{s^2}{\pi}\int_{4m_\pi^2}^\infty{
\frac{\rho(z) N(z)}{z^2
(z-s-i\epsilon)}dz }\label{eq:inversep}
\end{equation}

This equation expresses the pion vector form factor in terms of the N function
 of the elastic P-wave $\pi\pi$ scattering amplitude
 instead of the P-wave phase shift
$\delta_1$ as given by Eq. (\ref{eq:Omnes}). For the following purpose of calculation,  
Eq. (\ref{eq:inversep}) is more convenient. 
As mentioned above the function N(s) involves a dispersion integral
 of the discontinuity of the partial  wave amplitude across the left hand cut involving
the effects of the $I=0,1$  and  $2$ pair of pions exchanged in the crossed channels.
Because of the normalisation of the function D(0) = 1 and that
 the Weinberg low energy theorem is operative at low energy for the elastic 
$\pi\pi$ scattering amplitude, we can write the N function for the P-wave as:

\begin{equation}
N(s)=\frac{(s-4m_\pi^2)}{96\pi f_\pi^2}
 \{1+96\pi f_\pi^2\frac{(s-4m_\pi^2)}{\pi}\int_{0}^\infty 
\frac{ImN(-z)}{(z+4m_\pi^2)^2(z+s)}dz\} \label{eq:nf}
\end{equation}

Eq. (\ref{eq:inversep}) and this equation are still
 exact, apart from the assumption on the absence of the zeros.
 It provides an alternative solution to that given
by Eq. (\ref{eq:Omnes}) on how to
 construct the solution of the form factor problem when the dynamics 
of the strong interaction is known.
We now discuss an approximate scheme for Eq. (\ref{eq:inversep}). 
 To calculate the form factor V(s) in Eq. (\ref{eq:inversep}) we must
 know the function N(s) for $s>4m_\pi^2$. In this energy range, the contribution 
from the ImN(s) is usually small because the denominator in 
Eq. (\ref{eq:nf}) never vanishes. Hence 
we approximate N(s) by the first term on the R.H.S. of Eq. (\ref{eq:nf}),
 i. e. we neglect the contribution from ImN(s) or equivalently to use the Weinberg
low energy theorem for $\pi\pi$ scattering \cite{Weinberg1}.
 The approximation scheme is therefore to start initially with
 the Weinberg low energy theorem for the partial wave amplitude, then this result 
can be improved by correcting for the deviation of the Weinberg theorem 
at  higher energy. This procedure is reasonable because
 we use a twice subtracted dispersion relation in s 
 which emphasise the low energy contribution at small value of s and
 hence it is valid to approximate the $\pi \pi$ amplitude by the Weinberg expansion
 as a first approximation.
 
The form factor V(s) can now be written as:

                \begin{equation}
         V(s) = \frac{1} {1 -s/s_{R_1} - {1\over 96\pi^2f_\pi^2}\{(s-4m_\pi^2)
 H_{\pi\pi}({s}) + {2s/3}\}} \label{eq:vu}
\end{equation}

 Eq. (\ref{eq:vu}) can also be derived by using the Pad{\'e} approximant method:

\begin{equation}
V^{(0,1)} = \frac{V^{tree}}{1-{V^{1-loop}\over V^{tree}}} \label{eq:Pade}
        \end{equation}
          
where $V^{tree}$ refers to the tree amplitude which is equal to unity 
and $V^{1-loop}$ refers to the one loop amplitude, i.e. the last two
terms on the R.H.S. of Eq. (\ref{eq:pertv}).
The Pad{\'e} approximant method yields the same result as that
 given by Eq.(\ref{eq:vu}) which satisfies the elastic unitarity relation. 

We can justify  the use of the Pad{\'e} method to resum the one loop perturbation
series by looking at the imaginary part of the inverse of Eq. (\ref{eq:Pade}): it
is equal to $-\rho(z)N(z)$ with $N(z)$
 given by the tree amplitude    and hence containing no left hand cut, or $N(z)=
(s-4m_\pi^2)/(96\pi f_\pi^2)$.

The problem of $\pi \pi$ scattering, in the approximation where the 
contribution of the left hand cut could be neglected, was done a long time ago
by Brown and Goble \cite{Brown}. The more exact calculation with the contribution of
 the left hand cut and in the chiral limit
 was done by Lehmann \cite{Lehmann} where the real part of the logarithm terms
 due to the 
right and left hand cuts cancel each other out in the chiral limit. A more exact 
one loop perturbation calculation was done by \cite{GL1} and its unitarised form 
was given by \cite{Truong2}. In this last work, the calculated 
P-wave phase shifts agree very well with 
the experimental data from the $\pi\pi$ threshold to 1GeV. and differs little
 from those calculated by neglecting the left hand cut contribution.

Similarly, our calculation for the form factor using Eq. (\ref{eq:vu}) differs by at most 1-2\%
from the direct calculation of the form factor using Eq. (\ref{eq:Omnes}), with
 the P-wave  phase shifts given by the unitarised version of the P-wave $\pi \pi$ scattering 
\cite{Truong2}. Explicit results are given in reference \cite{Beldjoudi3}. Hence to the extent
 that the pion loops are taken into account, the left hand cut contribution to the N-function
of the P-wave amplitude can be neglected. 

 From another point of view, the problem of
treating the exchange of $\rho$ and $\sigma$ as particles with their full propagators,
in the framework of the generalised linear sigma model, was  
  previously examined in the reference \cite{Pham}. It was found that, in order
to have the validity of the KSRF relation, and treating the problem 
 at the tree level approximation, the effect
 of $\rho$ and $\sigma$ exchanged in the $t,u$ channels should cancel
 out approximately, when their masses satisfy
the relation $m_\sigma =\sqrt2 m_\rho$ which is not in contradiction with
the experimental fact. This problem should be further investigated.
 
 The above discussions give 
a justification for the approximation done by Brown and Goble \cite{Brown}. In
fact Eq. (\ref{eq:vu}) is exactly the inverse of the D-function which is the same as the
Omn{\`e}s function given by Eq. (\ref{eq:Omnes}).

In this section, we limit ourself to the approximation
 where the discontinuity of the left hand cut can be neglected in order to
make a connection of our calculation scheme with the perturbation series
and also with the large number of flavor $N_f$ expansion scheme. Objections
 can be raised against this approximation
 because of the large violation of the crossing symmetry which  we were taught back 
30 years ago to be an important property of the field theory. This opinion
is certainly correct, but calculations done at that period
 were without the constraints
of chiral symmetry, i.e. without having the low energy theorems which can be used to
write {\em subtracted dispersion relations}. Subtracted  dispersion
 relations are important because we want to suppress the high energy contributions 
which are difficult to calculate. With the use of the low energy
chiral theorems in dispersion relations, the crossing symmetry difficulties
are minimized as the scale of the physical problem at low energy is fixed by the low 
energy theorems. 

\vskip.15in
 { \bf c)  Deviation from the Elastic Unitarity Constraint}

As we shall show below, the inelastic contribution in
 the unitarity relation for the vector pion form factor
 due to the Kaon loops amounts to a few percent.
 The situation is quite different when we consider the contribution of the
  $\omega \pi$ and possibly the $\rho\pi\pi$ channels which are
 experimentally important. The calculation of
 this contribution to the pion form factor is quite complicated. We can derive a similar equation to the 
Muskelishvilli-Omn{{\`e}s} equation \cite{Pham2} but the phenomenological application
has considerable uncertainties \cite{Costa}.

Instead of the integral equation (\ref{eq:MO}), we have now:

\begin{equation}
V(s) = 1+\frac{s}{s_{R_1}} +\frac{s^2}{\pi}\int_{4m_\pi^2}^\infty{
\frac{V(z) f_1^*(z)+\sigma_i(z)}{z^2
(z-s-i\epsilon)}dz } \label{eq:MO1}
\end{equation}

where $f_1(s)=\frac{\eta e^{i\delta(s)}-1}{2i}$, $\eta$ being the inelastic factor,
 and $\sigma_i(s)$ being the inelastic contribution to the unitarity relation and differing 
from zero for s above the inelastic threshold $s_i$. The solution for this equation is \cite
{Pham2}:

\begin{equation}
V(s)=\frac{1}{D(s)} [1+\frac{1}{\pi}\int_{s_i}^\infty{\frac{2D(z) Re(\sigma_i e^{i\delta})dz}
{(1+\eta)e^{-i\delta}z(z-s-i\epsilon)}}] \label{eq:PT}
\end{equation}
We also have to require that the first derivative of $V(s)$ at s=0 to be $s_{R-1}$.

 Below and sufficiently far from $s_i$, e.g.
the $\omega \pi$ threshold, we can 
roughly parametrise the contribution of the second term on the R.H.S. of Eq. (\ref{eq:PT})
 as a polynomial $P_n(s)$ which is real for $s<s_i$ and is the polynomial ambiguity
in the solution of the Muskhelishvilli-Omn{\`e}s equation. Because
 of the normalisation $P_n(0)=1$, the introduction 
of this factor will not influence the Ward identity, but does influence the value of the
 r.m.s.value of the pion radius.  This type of approximation is reliable 
for energy below the inelastic threshold, but is erronous at the inelastic 
threshold and also at higher
 energy. 

        We shall fit the experimental data below with the simple expression $(1+\alpha s/s_\rho)$, 
where $\alpha$ is a constant in order to 
simulate phenomenologically the inelastic effect. The pion form factor can now be written as:

        \begin{equation}
         V(s) = \frac{1+\alpha s/s_\rho} {1 -s/s_{R_1} - {1\over 96\pi^2f_\pi^2}\{(s-4m_\pi^2)
 H_{\pi\pi}({s}) + {2s/3}\}} \label{eq:vu1}
\end{equation}

{\bf d) Phenomenological Applications}

\vskip .15in

       {\bf1)} Let us first calculate the form factor without using the phenomenological
 introduction of the inelastic effect in the unitarity relation.
 Eq. (\ref{eq:vu}) gives then the expression for the vector pion form factor.
 Because the vector r.m.s.
radius is positive, the vector form factor
 has a resonant character. Its width
 satisfies the KSRF relation \cite{KSFR}. The relation
 between the $\rho$ resonant  mass squared $s_\rho$
 and $s_{R_1}$ is:

\begin{equation}
s_{R_1} = \frac {s_\rho}{1-{1\over 96\pi^2f_\pi^2}\{ (s_\rho-4m_\pi^2)
 ReH_{\pi\pi}({s_\rho}) + 2s_\rho/ 3\}} \label{eq:sr1}
\end{equation}

In Eq.(\ref{eq:sr1}), we can either use $s_{R_1}$ or $s_\rho$ as an input parameter.
If the r.m.s. radius was used then one would predict the $\rho$ mass and 
 width which can be seen to satisfy the KSRF relation \cite{KSFR}: $\Gamma_\rho=
(s_\rho-4m_\pi^2)^{3/2}/
(96\pi f_\pi^2)$ where we have neglected the finite width correction \cite{Gounaris}.
 In terms of the ${g_{\rho\pi\pi}}$ coupling constant, we have: 

\begin{equation}
g_{\rho\pi\pi}^2 ={s_\rho\over {2 f_\pi^2}} \label{eq:KSFRP} 
\end {equation} 

where $g_{\rho\pi\pi}$ is defined at $s=s_\rho$. Using the experimental value 
$m_\rho=0.77 GeV$, the uncorrected width of the vector meson $\rho$ is 0.141 GeV.
 Including the finite width correction \cite{Gounaris}, the $\rho$-width, defined
 as $(\sqrt{s_\rho} \Gamma_\rho)^{-1} =
 cot\delta_1(s)^{'}\mid_{s=s_\rho}$, where the sign $ ^{'}$ denotes the first derivative
of $cot\delta_1$ with respect to s. 
The KSRF relation, with the finite width correction, yields $\Gamma_\rho=155 MeV$
which is very near to the experimental value of $151.5\pm 1.5 MeV$.

If one used $s_\rho$ as an input parameter, then one would 
predict the $\rho$ width by the KSRF
 relation \cite{KSFR}, and the r.m.s. radius of the pion.
For a number of reasons which will become clear later, we use the $\rho$ mass as an input.
 The calculated r.m.s. radius is
$<r_V^2>=0.40 \pm 0.01 fm^2 $ compared with the experimental value $<r_V^2>=
 0.439\pm 0.008 fm^2$ \cite{rms}. The calculated value is therefore too low.
This difficulty is  related to the fact that the calculated pion form factor at the
 $\rho$ peak
 is also too low, as can be seen in Fig. (1) (see below). This result is expected, because we neglected
 the inelastic contributions in the unitarity relation which does not change 
the imaginary part of the pion form factor but does contribute to
 its real part as explained above. A much better agreement with the experimental data,
 both at the $\rho$ peak and also for the pion r.m.s. radius can be obtained by
 phenomenologically introducing the inelastic effect.

        The phase and modulus squared of the pion form factor calculated by the CPTH,
Eq. (\ref{eq:pertv}), and  by UCPTH, Eqs.(\ref{eq:vu},\ref{eq:Pade}),
 using $<r_V^2>= 0.40 \pm 0.01 fm^2 $
 are shown on Figs. (1) and (2). Although they both
agree with each other at low very momentum transfer, due to the dominance of the 
r.m.s radius term, the 
higher energy  experiment data are
 much better represented by the UCPTH calculation than those of the CPTH calculation.
Experiences with analytic functions show that a small difference between
 two analytic functions in one region can be greatly amplified in another region.
This explains why the form factors calculated in both CPTH and UCPTH  
have the same value and radius at s=0, but have completely different behavior
 around the $\rho$ resonant region. The UCPTH calculation has the $\rho$ resonance
 character while
 that from the one loop CPTH does not.

The phase theorem of the pion form factor \cite{Watson} can be used to test 
whether the unitarity relation is satisfied or not. From Fig.( 2), for $s >0.2 GeV^2$ 
the CPTH phase is clearly much less than those obtained
  experimental data reflecting that the unitarity relation is badly violated
in this approach. The UCPTH pion form factor phase is, on the other hand, in very good
agreement with the data from the $2 \pi$ threshold to $1 GeV^2$. This reflects
 that the UCPTH approach satisfies the unitarity relation in this region. We
 conclude that the unitarity relation plays a crucial role in the low energy 
calculation of the pion form factor and also of the $\pi \pi$ scattering. We shall
 show elsewhere some difficulties of calculating the process $\pi \to \gamma \gamma^*$
encountered in the CPTH approach can also be removed by using the
 UCPTH method \cite{Truong6}.

\vskip 0.1 in
{\bf e) Inadequacy of CP¬TH calculation for Vector Form Factor.}

Let us now examine Eq. (\ref{eq:pertv}). As noted above,
  the last term in Eq. (\ref{eq:pertv}) has
 a different low s behavior
for the real and imaginary parts: at small s, its real part behaves as $s^2$ while 
its imaginary parts, in the chiral limit, behaves as s. This mismatch between the
small s behavior of the real and imaginary parts is due 
  to the perturbative approach which results in a violation of the phase theorem.
 In order to restore this theorem, even at low energy, the next order term
in the imaginary part has to be included in the perturbation calculation as 
pointed out in ref. \cite{Truong1} and will be discussed later
 in the scalar form factor calculation.
 This problem does  not exist in UCPTH approach as can be seen from Eq. (\ref{eq:vu}).

From the discussion of the previous paragraph, we are led to the study of the 
O($s^2$) term in Eq. (\ref{eq:pertv}) and that in Eq. (\ref{eq:vu}).
 Although they are numerically very small
 at low values of s, because we are interested
in extending the region of the validity of the perturbation theory 
to near the $\rho$ resonant region, they must be taken into account in this energy
range. 
Hence we are led to examine the consistency of the CPTH approach compared with
the UCPTH calculation or the VMD model which we shall show later to be equivalent. For this purpose, not only
 we want to compare the experimental data on the modulus of the pion form factor
 but we also want to use the phase theorem which is model independent, to compare
 the calculated phases with those obtained from the experimental data.
 
 From Eq. (\ref{eq:vu}), it is seen that the perturbative
 approach of Eq. (\ref{eq:pertv}) leaves out, at low energy, a 
term $(s/s_{R_1})^2$ which should be compared
with the real part of the last term in Eq. (\ref{eq:pertv})
because they are both of the same order
in s. The perturbative approach is only justified
if the last term is much larger than the former.
 Let us now define the parameter
$r_V$, defined as the ratio of the real
 part of the last term on the R.H.S. of Eq. (\ref{eq:pertv}) over $(s/s_{R_1})^2$ :

\begin{equation}
r_V = \frac{{1\over 96\pi^2f_\pi^2}  ((s-4m_\pi^2)
 ReH_{\pi\pi}({s}) + {2s\over 3} )}{(s/s_{R_1})^2}
\end{equation} \label{eq:rv}

In Fig.(3), the ratio $r_V$ is plotted as a function of s.
 It is seen that, over a wide range of values of s, including small values of s , $r_V$ is much 
less than unity instead of being much larger than unity in order 
to guarantee the validity of the perturbation theory. As we shall show below
the situation is very different in the calculation of the scalar form factor.

We have discussed so far, the CPTH one-loop calculation of the pion form factor. It 
is not difficult to calculate the two-loop contribution to the vector pion form factor as
was previously done numerically by Gaisser and Mei$\beta$ner \cite{Gasser3} and more recently
 analytically by Colangelo et al. \cite{Colangelo}. The main point is that 
 the two loop calculation requires an extra  subtraction in the dispersion relation
 and hence it is necessary to introduce an extra parameter, the second derivative of
 the pion form factor at the origine.
 These authors fix this parameter by taking 
the VMD model  which is  $(s/s_\rho)^2$. What one calculates in the CPTH approach 
 is the contribution of the
 $s^3$ terms, modulo of the logarithm. One  can then show the result of the calculated $O(s^3)$ term  
 is still smaller than the $(s/s_\rho)^3$ terms coming from
 the VMD model, although the corresponding
$r_V$ term is, on the average, a factor of 2-3 larger than the value of 
$r_V$ of the one loop-calculation. 
        
Although we have not carried out the calculation of the 3-loop and higher loop amplitudes,
 we suspect the 
same difficulty also occurs. This  unsatisfactory situation of the 1-loop, 2-loop and possibly
 higher loop calculations, is due to the use of 
the perturbation theory to study nonperturbative phenomena. One would
 have 
to increase the $\rho$ mass to be larger than 1.5 GeV in order to make
 accuracy of the vector form factor calculation comparable to the
 scalar case(see below).

It is so much simpler and more transparent to use Eqs. (\ref{eq:vu},\ref{eq:Pade})
 which is even better than the VMD model as a starting point. One then improves it by taking
 into account of the 2-loop graphs which are neglected in the calculation.
 A better expression, which takes into 
account of the inelastic effect yielding a correct r.m.s. radius and the height of the
 $\rho$ peak Eq. (\ref{eq:vu1}), is even better than
 Eqs. (\ref{eq:vu},\ref{eq:Pade}) for this purpose. If the major goal of physics was to describe 
experimental data by a few parameters, because CPTH had to rely on the VMD model
to get the necessary counter terms, it would become a rather special way of doing physics. 

       We would like to ask whether by including 
higher order terms in the CPTH approach that one could generate the Breit-Wigner form.
 We have a strong doubt about this possibility. Our feeling is based on
 the experimental fact that there is very little inelastic effect below 
1 GeV in the form factor so that higher loop effect cannot be important in the
$\rho$ resonance region.
 
 In Fig. (4) we plot the modulus of the pion form factor calculated by the
UCPTH compared with 
 the sum of the first four terms of the Taylor series expansion
 of the UCPTH calculation.
 It is seen that, at low energy, the first four 
terms are a good approximation for the experimental data but we cannot
 make the the Breit Wigner form i.e. to make the high energy curve to
 turn down. This illustrates the problem of the non convergence of the 
perturbation series for the amplitude; there is, however, no difficulty in generating
the Breit Wigner form by expanding the inverse amplitude as a Taylor's
 series of momenta. Our remark here is a relativistic
 generalisation of the Bethe's "effective range theory"
to handle the resonance and bound state in the two nucleon
 system \cite{Bethe}.
\vskip.15in
{\bf2)} We want now to improve our calculation by taking into account of the inelastic effect.
 As explained above, the simplest parametrisation can be done by setting $P_n(s)=
 1+\alpha s/s_\rho$ and is given by Eq. (\ref{eq:vu1}). 

The good fit to the data can be obtained with $\alpha=0.15$ with
$\sqrt{s_\rho}=O.77 GeV$. The  vector pion r.m.s. radius squared 
is now  $<r_V^2>= 0.46 \pm 0.01 fm^2 $ compared with the experimental value $<r_V^2>=
 0.439\pm .008 fm^2$\cite{rms}.  A slightly better fit to the data is obtained with $\alpha=0.13$
 but the $\rho$ mass is now changed to 0.773 GeV;  the vector pion r.m.s. radius squared 
is now  $<r_V^2>= 0.45 \pm 0.01 fm^2 $ which is quite good.

Before leaving this section let us write down the dispersion relation for
 the form factor when the r.m.s. radius is not used as a subtraction constant, but
a low energy measured form factor at $s=-s_0$ is used as a subtraction constant. Let us consider
 a measured space-like form factor at $s=-s_0$ (the time-like form factor
can straightforwardly be written down). Instead of Eq. (\ref{eq:MO}), we now have:
\begin{equation}
V(s)=V(0) +(V(0)-V(-s_0))\frac{s}{s_0}+\frac{s(s+s_0)}{\pi}
\int_{4m_\pi^2}^\infty{
\frac{V(z)  e^{-i\delta_1(z)}sin\delta_1(z)}{z(z+s_0)
(z-s-i\epsilon)}dz } \label{eq:mod}
\end{equation}

The one-loop CPTH result can be obtained perturbatively from this equation by setting 
in the integrand $V(z)=1$ and $f_1^{tree}(s) = 
 e^{i\delta_1(s)}sin\delta_1(s)/\rho(s) = s/(96\pi f_\pi^2)$. 
 Because we use the
 input as the measured form factor at $s=-s_0$, we cannot discuss the obtained results
as an expansion in a power series in s.

Let us finally discuss the ghost problem in the Pad{\'e}, inverse or the N/D methods. 
 Eqs.(\ref{eq:vu},\ref{eq:Pade}) develop a pole at $s_g=-3.8\cdot10^5 GeV^2$
 which is a ghost and should not be there. It has to be eliminated from these equations by 
multiplying them with the factor $(1-s/s_g)$ which does not effect the normalisation 
of the form factor at $s=0$, but change the value of the pion form factor by an amount
 $s/s_g$ which is completely negligible at small s.
 The phase theorem is also
 unaffected by this factor. We show here that the presence of a ghost due to the unitarisation 
procedure can be tolerated, as long as it is sufficiently far from the 
physical region of interest.

\section{Calculation of the Scalar Form Factor}

Similarly to the calculation of the vector pion form factor, we can now calculate the 
isoscalar scalar form factor S(s) defined as:
\begin{equation}
 <\pi^{a}(p_1)\pi^{b}(p_2)\mid
m(\bar{u}u+\bar{d}d)\mid0> = m_\pi^2\delta^{ab} S(s)
\end{equation}
where $s=(p_1+p_2)^2$. After introducing the scalar r.m.s. radius,
 using the definition:  $S ^{'}(0) =\frac{1}{6} <r_S^2> = 1/s_{R_2}$ to
eliminate the scale dependent $\mu$, the one-loop perturbative result 
for the scalar form factor is:
        \begin{equation}
                 S^{pert.}(s) = 1 +\frac{s}{s_{R_2}}\ + {1\over 16\pi^2f_\pi^2} ((s-m_\pi^2/2)
 H_{\pi\pi}({s}) + {s\over 12} ) \label{eq:perts}
\end{equation}

After unitarisation, the UCPTH result for the scalar form factor is:

\begin{equation}
         S(s) = \frac{1}{1 -\frac{s}{s_{R_2}}\ - {1\over 16\pi^2f_\pi^2} ((s-m_\pi^2/2)
 H_{\pi\pi}({s}) + {s\over 12} )} \label{eq:su}
\end{equation}

Similarly to the definition of $r_V$, we can define $r_S$,
 the ratio of the $O(s^2)$ of the CPTH calculation  and $(s/s_{R_2})^2$:

\begin{equation}
r_S = \frac{{1\over 16\pi^2f_\pi^2}((s-m_\pi^2/2)
ReH_{\pi\pi}(s) + {s\over 12} )}{(s/s_{R_2})^2} \label{eq:rs}
\end{equation}

In Fig. (3), the ratio $r_S$ is plotted against the energy squared. It is seen
that $r_S$ is larger than unity, although not much larger on the average;
  hence, we are assured about the approximate validity of the perturbation theory. 
The reason for the difference with the vector form factor calculation is due to
the coefficient of the chiral logarithm term being larger by a factor 6 in
the scalar form factor. This fact explains the partial success of the one loop CPTH
 calculation in the S-wave $\pi\pi$ scattering ( this was done with the  prescription
 that the phase shift
 is  proportional to the real part of the partial wave amplitude
 in $\pi\pi$ elastic scattering \cite{GL1}. Had the calculation been done
with the definition $tan\phi= Imf/Ref$, the result would be very different). 

It should be remarked
that in the CPTH approach, the calculated phase of the form factor is not
 the same as the phase shift $\delta$ because the unitarity relation is not
satisfied in the perturbative approach. In fact, if we calculated
 the phase of the form factor in the CPTH approach, Eq. (\ref{eq:perts}),
 we would find out that:

\begin{equation}
tan(\phi_{CPTH})=\frac{\rho(s)(s-m_\pi^{2}/2)}{16\pi f_\pi^2 \{1 +s/s_{R_2}
 + {1\over 16\pi^2f_\pi^2} [(s-m_\pi^2/2)
Re H_{\pi\pi}({s}) + {s\over 12}] \}} \label{eq:phip}
\end{equation}
to be compared with the phase calculated by the UCPTH method, Eq. (\ref{eq:su}):

\begin{equation}
tan(\phi_{UCPTH})=\frac{\rho(s)(s-m_\pi^{2}/2)}{16\pi f_\pi^2\{1 -s/s_{R_2}
 - {1\over 16\pi^2f_\pi^2} [(s-m_\pi^2/2) 
Re H_{\pi\pi}({s}) + {s\over 12}] \}} \label{eq:phiu}
\end{equation}

In Fig.(5) the dashed curve is the result of the standard
one loop CPTH using the same parameter as that of UCPTH
Eq. (\ref{eq:pertv}). The phases of the scalar form factor calculated by CPTH and UCPTH,
$\phi_{CPTH}$ and $\phi_{CPTH}$ are 
plotted against the energy squared. It is seen the phase of the
form factor calculated by CPTH is
 in a much better agreement with the data than the vector case; the agreement
between the UCPTH calculation and the experimental data is, however, better than
the {\em phase} of the one-loop CPTH calculation. It should be recalled that the 
phase shift calculated by the CPTH method was done by making the prescription that 
it is proportional to the real part of the strong S-wave $\pi\pi$ 
partial wave and agrees well with the experimental data.
The difference between the CPTH {\em phase} of the form factor and the phase shift is
therefore a measure of the violation of the unitarity relation.

It is seen that these two phases are different even at the threshold.
Let us compute the limit $s\rightarrow 4m_\pi^2$ for $tan\phi/\rho(s)$.
 From these two equations and use the 
definition for the scattering length $a$, the limit $s \rightarrow 4m_\pi^2$ of 
$tan\phi=(1/2) \sqrt{s-4m_\pi^2} a$, we  get:

\begin{equation}
a_{CPTH}=\frac{7m_\pi}{32\pi f_\pi^2}(1+{4m_\pi^2\over s_{R_2}}+{11m_\pi^2\over
{24\pi^2 f_\pi^2}})^{-1} 
\end{equation} \label{eq:acpth}
and
\begin{equation}
a_{UCPTH}=\frac{7m_\pi}{32\pi f_\pi^2}(1-{4m_\pi^2\over s_{R_2}}-{11m_\pi^2\over
{24\pi^2 f_\pi^2}})^{-1} \label{eq:ucpth} 
\end{equation} 

For $s_{R_2}=0.56 GeV^2$ which agrees with the experimental scalar
radius $<r_s^2>=0.41fm^2$, we 
have: $a_{CPTH}=0.123m_\pi^{-1}$ and  $a_{UCPTH}=0.23m_\pi^{-1}$.
These values are to be compared with the direct CPTH calculation of $\pi\pi$ scattering
 $a=0.20m_\pi^{-1}$ \cite{GL1} and the same process using
 UCPTH,   $a=(0.26\pm0.02)m_\pi^{-1}$ \cite{Truong2}.

It is clear that the phase theorem for the form factor is violated in 
the CPTH method. How can we rectify this situation for CPTH method? As we
 mentioned above, there is a mismatch between the real and imaginary part
of our calculation.
In the perturbative calculation, the real part of the form factor is calculated to 
the second order in s while its imaginary part is only calculated to the first order
in s. 
In order to compute correctly, in the CPTH approach,
the phase $\phi$ of the form factor given by $tan\phi=ImS/ReS$,
 we must also calculate 
the imaginary part of the form factor to the second order in s, i.e. the two loop
 order for its imaginary part. More precisely, the perturbative unitarity relation
for ImS including the two loop contribution is:

\begin{equation}
ImS(s)=\rho(s)\{S_1(s)f_1(s)^*+S_2(s)f_1^*(s)+S_1(s)f_2^*(s)\}
\end{equation} \label{ims}

where $S_1$ and $S_2$ are, respectively, the first and the remaining terms
on the right-hand side of Eq. (\ref{eq:perts}); they represent the tree and one 
loop amplitudes. Similarly $f_1$ and $f_2$ are the correponding elastic $\pi\pi$
scattering amplitudes. The calculated scattering length is now :

\begin{equation}
a={7m_\pi\over 32\pi f_\pi^2}\{1+{Ref_2(4m_\pi^2)\over f_1(4m_\pi^2)}(1+{4m_\pi^2\over 
s_{R_2}}+
\frac{11m_\pi^2}{24\pi^2 f_\pi^2})^{-1}\}
\end{equation} 

Using the value $Ref_2(4m_\pi^2)/Ref_1(4m_\pi^2) \simeq 0.25 $ from the one loop 
calculation for the I=0  $\pi\pi$ scattering amplitude \cite{GL1,Truong2}, this Eq.
 yields
$a=0.195m_\pi^{-1}$. This value is very close to the scattering length
 $a=0.20m_\pi^{-1}$
 obtained by CPTH method \cite{GL1} 
for the elastic $\pi\pi$ scattering using the prescription that the phase shift
$\delta$ is equal to real part of the calculated partial wave. 

We see that by correcting for the low energy mismatch of the real 
 and the imaginary parts
we can get a reasonable agreement with the phase theorem. A small discrepancy between 
the new value of the scattering length $0.195 m_\pi^{-1}$,
 with that obtained from UCPTH
 approach $0.23m_\pi^{-1}$, is due 
to the fact that $r_S$ given by Eq. (\ref{eq:rs}) is not much larger than unity.

For a more complete 2 loop calculation of the scalar pion form factor, see the recent work of 
 Gasser and Mei$\beta$ner and Dognohue et. al. \cite{Gasser3},\cite{Gasser4}. It should be 
remarked that the sum of the last 2 terms on the R.H.S. of Eq. (\ref{ims}) is 
 not necessarily real, one must take their real part in the calculation.

Eq. (\ref{eq:su}) develops a pole at $s_g = -2.7 GeV^2$ which is a ghost and has to be removed 
by multiplying the R.H.S. of this Eq. a factor $(1-s/s_g)$ which does not
 affect the phase theorem, but does change the modulus of the form factor. Unlike the vector 
pion form factor, this ghost is  near to the physical
 region and has to be taken into account. At $s=m_K^2$,
 where $m_K$ is the Kaon mass, the correction enhances 
the modulus of the scalar pion form factor by 10{\%}.

To end this section let us remark that in a recent study,
 the role of the left hand cut for the scalar form factor
 has recently been studied in details by Boglione and Pennington\cite{Pennington} in connection 
with the criticism of the recent work of Dobado and Pelaez \cite{Hannah}.
 They found that this method is an unreliable 
way of determining the chiral parameters but  a good fit to the data 
can be done with        an appropriate adjustment of the subtraction constant. 
 We agree with this statement, but would like 
to point out that the situation for the vector pion form factor is different as discussed above.

\section {UCPTH and Vector Meson Dominance}

 In the zero width approximation and at the tree level,
 the vector pion form factor can be written as:

\begin{equation}
  V^{0}(s) = \frac{f_\rho^0 g_{\rho\pi\pi}^0}{s_\rho - s}  
\end{equation}

where $f_\rho^0$ and $g_{\rho\pi\pi}^0$ are, respectively, the photon-$\rho$ coupling
( multiplying with e) and the strong $\rho\pi\pi$ coupling with
 $f_{\rho}^0 g_{\rho\pi\pi }^0/m_\rho^2=1$. At the tree level, these coupling 
constants should be independent of s, but loop corrections to the 2 and 3 point functions
 can introduce 
the s dependence. We now want to make a finite width correction to this Eq. by introducing
the self energy correction for the $\rho$ propagator. Summing the geometric
 series for the $\rho$ self
energy and performing the mass and wave-function renormalisations, we have:

\begin{equation}
 V(s) =\frac{f_\rho (s) g_{\rho\pi\pi }(s)}{s_\rho - s-\pi_\rho (s)}
\end{equation} \label{eq:v}

with 

\begin{equation}
Re\pi_\rho(s)= \frac{(s-s_\rho)^2}{\pi} P\int_{4m_\pi^2}^{\infty}
\frac{Im\pi_\rho(z)-Im\pi_\rho(s_{\rho})-(z-s_{\rho})Im\pi^{'}_\rho(s_{\rho})}
{(z-s_{\rho})^2  (z-s) } dz \label{eq:pi}
\end{equation}

and

\begin{equation}
Im\pi_\rho(s)={1\over 48\pi}\frac{(s-4m_{\pi}^2)^{3/2}}{\sqrt{s}}g_{\rho\pi\pi}^2
(s) \label{eq:impi}
\end{equation}

where P stands for the principal part integration.
The R.H.S of Eq. (\ref{eq:pi}) could be straightforwardly expressed in terms of 
the function $H_{\pi\pi}(s)$ and its derivatives,
 if $g_{\rho\pi\pi}(s)$ was a constant. 
In the usual VMD model,  $g_{\rho\pi\pi}(s)$ and $f_\rho(s)$ are constants 
 hence $g_{\rho\pi\pi}f_\rho/(s_\rho-\pi(0)) =1$. If we neglected the finite width 
correction, we would have  $s_{R_1}=s_\rho$ and hence the KSRF relation becomes
 $g_{\rho\pi\pi}^2 = s_{R_{1}}/(2 f_{\pi}^2)$. Let us now use a theorem stating that 
two real analytic functions, having the same discontinuity, can only differ by 
a real polynomial. By comparing Eqs. (\ref{eq:pi}), (\ref{eq:impi}) with Eq. (\ref{eq:vu})
 we see that they 
have the same imaginary part and satisfy the same boundary condition and high energy
behavior, hence they must be identical. In sum, we have the equivalence between
the UCPTH and the VMD.

We could improve the Vector Meson Dominance result by introducing the inelastic effect
in the vertex correction, e.g. the $\rho \rightarrow\omega\pi$  to 2$\pi$ state. This correction can be 
phenomenological represented by a real polynomial in s, and in order to fit the
experimental r.m.s radius, 
we could choose this polynomial as $(1+0.15s/s_\rho)$ which yields the same result
 as the calculation of the pion form factor with the inelastic contribution, Eq. (\ref{eq:vu1}).

 \section{Generalisation to SU(3)}

It is straightforward to generalise the CPTH and UCPTH to SU(3). We have
 to calculate in addition, the $K\bar{K}$ contribution to the pion form factor. Similarly
we calculate the
 $\pi K$ vector form factors with the $\pi K$ and $\eta K$ contributions and also the 
$K\bar{K}$ and $\eta K$ form factors. Because the scalar $\eta K$ form factor is negligible 
due to the small difference  of the K and $\eta$ masses; the scalar $\pi K$ scalar
 form factor is previously well calculated, we refer the readers to the original 
calculation \cite{Beldjoudi1}.
        Because we later want to calculate the $\tau$ decay rate, let us now calculate 
the charge current (isovector) matrix element which we denote now by $V_{ij}$, 
where i and j denote the pseudoscalars. The normalisation of $V_{ij}$ at 
zero momentum transfer is given by the Ademollo-Gatto theorem \cite{Gatto}.
        Using UCPTH, the $\pi^+\pi^0$ form factor is given by:

 \begin{equation}
                V_{\pi^+\pi^0}(s) =\sqrt{2}\{1 -\frac{s}{s_{R_1}}\ - {1\over 96\pi^2f_\pi^2}
 [(s-4m_\pi^2)
 H_{\pi\pi}(s) + {2s\over 3}+{1\over 2}((s-4m_K^2 ) H_{K\bar{K}}(s) +
 {2s\over 3})]\}^{-1}  \label{eq:vpipi}
 \end{equation}

where the subscripts in the function H refer to the intermediate states 
 defining this function. Similarly we calculate the $K^+K^0$ 
isovector vector form factor:

 \begin{equation}
                V_{K^+\bar{K}^0}(s) =\{1 -\frac{s}{s_{R_1}}\ - {1\over 96\pi^2f_\pi^2}
 [{1\over 2}((s-4m_\pi^2)
 H_{\pi\pi}(s) + {2s\over 3})+((s-4m_K^2 ) H_{K\bar{K}}(s) +
 {2s\over 3})]\}^{-1}\label{eq:vkk} 
\end{equation}

It should be noticed at this point that in the VMD calculation, the $V_{\pi^+\pi^0}(s)$
and $V_{K^+\bar{K^0}}(s)$ should be equal and is given by the expression for
$V_{\pi^+\pi^0}(s)$. 
 
The problem of $K\pi$ form factor is discussed in
some details in reference \cite{Beldjoudi1}. We quote
 their results on the calculation of
the contribution of $K\pi$ intermediate state and add to it the contribution of 
the $K\eta$ intermediate state (or loop). To simplify our writing, let us define 
the following function for the unequal mass case:

\begin{equation}
\bar{H}_{ij}(s,m_i^2,m_j^2) = s^2 \int_{(m_i+m_j)^2}^{\infty}
\frac{\lambda^{3/2}(z,m_i^2,m_j^2)/z^{3/2}}{z^2 (z-s-i\epsilon)}dz  \label{eq:hbar}
\end{equation}

where $\lambda(s,m_i^2,m_j^2) = (s-(m_i+m_j)^2)(s-(m_i-m_j)^2)$.
The function $\bar{H}_{ij}(s,m_i^2,m_j^2)$ can be expressed in terms of the logarithm 
function as was done in the references\cite {GL2,Beldjoudi1}.
 In term of this function, we can write
down now the expression for the
 $K^0\pi^+$ and $K^+\pi^0$ vector form factors:

 \begin{equation}
                V_{K^0\pi^+}(s) = \{1-\frac{s}{s_{R_3}} -{1\over 128\pi^2f_\pi^2}
 (\bar{H}_{\pi K}(s,m_\pi^2,m_K^2)+
\bar{H}_{\eta K}(s,m_\eta^2,m_K^2))\}^{-1}\label{eq:vkpi}
\end{equation}

\begin{equation}
                V_{K^{+}\pi^0}(s) ={1\over \sqrt{2}} \{1-\frac{s}{s_{R_3}} -
{1\over 128\pi^2f_\pi^2}(\bar{H}_{\pi K}(s,m_\pi^2,m_K^2)+
\bar{H}_{\eta K}(s,m_\eta^2,m_K^2))\}^{-1}
\label{eq:vvkpi}
\end{equation}

and the $K\eta$ form factor is given by:

\begin{equation}
                V_{K^{+}\eta}(s) =\sqrt{3\over 2} \{1-\frac{s}{s_{R_3}} -{1\over 128\pi^2f_\pi^2}
 (\bar{H}_{\pi K}(s,m_\pi^2,m_K^2)+
\bar{H}_{\eta K}(s,m_\eta^2,m_K^2))\}^{-1}\label{eq:vketa}
\end{equation}

Strictly speaking, all above equations are derived not by the inverse
amplitude method but rather by the Pad{\'e} approximant method. 
It should also be mentioned that the  $K^{+}\bar{K}^0$ and the $K^{+}\eta$ form factors 
cannot be  calculated at their threshold values by CPTH technique because
 their thresholds are above the resonant masses.

The  vector pion r.m.s. radius squared is calculated to be $<r_V^2>= 0.40 \pm 0.01 fm^2 $ compared with the experimental value $<r_V^2>=
 0.439\pm .008 fm^2$ \cite{rms}.
The vector $K\pi$ r.m.s. radius squared is calculated to be
 $<r_V^2>= 0.27 \pm 0.008 fm^2 $ compared with the experimental value $<r_V^2>=
 0.34\pm .03 fm^2$\cite{Montanet}. The agreement between the
 theory and the experimental data is not satisfactory (see below).

        Including the inelastic effect due to the $K^{+}\bar{K^0}$ intermediate state in the
 pion form factor calculation, changes little the numerical result. In a more
 elaborate calculation \cite{Beldjoudi3} where the $\pi\pi$ phase shifts were calculated 
first which includes the left hand cut contribution to the partial wave
strong $\pi\pi$ amplitudes and then compute the vector pion form factor
using Eq. (\ref{eq:Omnes}), there was very little change from the
 result of the present calculation.
The contribution of the
 $ \omega\pi$ inelastic effect in the unitarity equation for the form factor is,
however important \cite{Costa}.

The effect of the left hand cut in the  P-wave elastic amplitude is, however,
 more important in the Vector $K\pi$ form factor
 calculation \cite{Beldjoudi1,Beldjoudi3}. 

\section{Application to Tau Decays}

\vskip .15in
Obviously CPTH cannot be used for calculations in $\tau$ decays except
 for a very small region of the phase space where the one-loop calculation
is valid \cite{Colangelo}. In this region of the phase space, the CPTH results
are the same as those obtained by the UCPTH,   but the latter method
has a much wider range of validity.
 Because UCPTH can handle resonance,
we can use this method to calculate the $\tau$ hadronic decays.
Let us now calculate the decay of the lepton $\tau$ into 2 pseudoscalars i and j.
Let us define the ratio $R_{ij}$ of the rate of $\tau \rightarrow P_i P_j\nu$ 
to that of
$\tau\rightarrow e\nu\bar{\nu}$. Taking only into account of the vector form factor
contribution,
we have:

\begin{equation}
R_{ij} = {1\over {4m_\tau^2}}(\begin{array}{c}
cos^2\theta_c \\
sin^2\theta_c \\ \end{array})\int_{s_t}^{M^2}\frac{\lambda(s,m_i^2,m_j^2)^{3/2}}{s^3}
(1+2s/m_\tau^2)(1-s/m_\tau^2)^2 \mid V_{ij}(s)\mid^2 ds \label{eq:rij}
\end{equation}

where $\theta_c$ is the Cabbibo angle with $sin^2\theta_c = (0.222)^2$,
 $s_t$ is the square of the threshold energy  and M is the $\tau$ lepton mass.
 Let us use the following notation $R_{ij}(ab)$
to denote the ratio R defined in Eq. (\ref{eq:rij})
 with intermediate states a and b and
$R_{\pi K}$ means the sum of the decay rate into $\pi^{+} K^0$ and $\pi^0 K^{+}$.
 Using Eqs. (\ref{eq:vpipi}-\ref{eq:vketa}), we have:
\vskip.15in
\begin{eqnarray}
R_{\pi^{+} \pi^0}(\pi\pi) = 1.05 & R_{\pi^{+} \pi^0}(\pi\pi, K\bar{K})=1.03 
& R_{\pi^{+} \pi^0}\mid_{exp} = 1.38\pm .02 \\
R_{K^{+} \bar{K}^0}(\pi\pi) =0.0042    & R_{K^{+} \bar{K}^0}(\pi\pi,K \bar{K})=0.0031  &
 R_{K^{+} \bar{K}^0}\mid_{exp}= .0075\pm .002 \\
R_{\pi K}(\pi K)=0.050 & R_{\pi K}(\pi K,\eta K)=0.048          
 & R_{\pi K}\mid_{exp} =0.065\pm .008 \\
R_{\eta K}(\pi K) = 5.1.10^{-4}  & R_{\eta K}(\pi K,\eta K)=4.1. 10^{-4}
   &R_{\eta K}\mid_{exp}= (1.3\pm .04) 10^{-3} 
\end{eqnarray}

The experimental data are taken from the recent paper of the CLEO group\cite{Cleo} 
and the Particle Data Group \cite{Montanet}.

\vskip.15in
It is seen that our calculation for the $\rho$ and $K^*$ decays are too small by 
35\% compared with the experimental rates. This is expected because we neglect the
 inelastic contribution and 
essentially use the threshold parameters,
the r.m.s. radii as inputs to extrapolate to the resonance region which is far away.   

Similarly the $K^+\bar{K}^0$ and $K^+\eta$ decays are too low by a factor of 2.
(The finite width correction is very small for the  $K^*$: the KSRF relation gives 
$\Gamma_{K^*}$ = 55.2 MeV and the finite width correction with $\pi K$ loop gives 
55.6 MeV and with $\pi K$ and $\eta K$ loops give 53.4 MeV).
The scalar form factor contribution to the above decay rates are small. Their
 maximum contribution is in$R_{\pi K}$ which amounts only to 3-4\% of the calculated 
rate \cite{Beldjoudi1}.
\vskip .15in

 If we are willing to introduce more
 parameters in our calculation 
 in order to fit the  data on the top of the resonances and want only to predict the 
r.m.s. radii and the inelastic form factors $K^+\bar{K}^0$ and $K^+\eta$,  
we can certainly do a better job. This can be done by using the polynomial
in Eq. (\ref{eq:Omnes}) as explained above.

Because on the top of the $\rho$ resonance the peak value of the
 form factor is too low by 30\%,
 we can set $P_n(s)= (1+0.15 s/s_\rho)$ in order to have the correct
 peak value, and hence we multiply the right hand
side of Eqs. (\ref{eq:vpipi}-\ref{eq:vketa}) by this factor.

The  vector pion r.m.s. radius squared is now  $<r_V^2>= 0.46 \pm 0.01 fm^2 $ compared with the experimental value $<r_V^2>=
 0.439\pm .008 fm^2$\cite{rms}.
The vector $K\pi$ r.m.s. radius squared is now
 $<r_V^2>= 0.31 \pm 0.008 fm^2 $ compared with the experimental value $<r_V^2>=
 0.34\pm .03 fm^2$ \cite{Montanet}. There is now a good agreement with the data.

The $\tau$ decay rates are now:
\vskip .15in
\begin{eqnarray}
R_{\pi^+\pi^0}(\pi\pi) = 1.41 &R_{\pi^+\pi^0}(\pi\pi, K\bar{K} )=1.38 
& R_{\pi^+\pi^0}\mid_{exp} = 1.38\pm.02 \\
R_{K^+\bar{K}^0}(\pi\pi) = 0.067   & R_{K^+\bar{K}^0}(\pi\pi,K\bar{K})=0.064
  & R_{K^+\bar{K}^0}\mid_{exp}=  .075\pm .02 \\
R_{\pi K}(\pi K)=0.067 & R_{\pi K}(\pi K,\eta K)=0.064 
 & R_{\pi K}\mid_{exp} = 0.065\pm .008 \\
R_{\eta K}(\pi K) = 8.9. 10^{-4}  & R_{\eta K}(\pi K,\eta K)=7.2 .10^{-4}
   &R_{\eta K}\mid_{exp}=(1.3\pm .04) 10^{-3} 
\end{eqnarray}
\vskip .15in

 It is seen
 that the agreement with the data is much better which is not surprising because 
we do not have such a long range of energy to extrapolate, from the resonant
 energy to the $\eta K$ and $ K\bar{K}$ energy available in the $\tau$ decay.
These results are changed slightly when we change the $\rho$ mass to 0.773 GeV and the 
correction factor to $(1+0.13s/s_\rho)$.

\section{\bf Conclusion}

We have presented here the  study of the form factor problem using the UCPTH approach.
The main feature of this method is that, at low energy, it coincides with that derives
from the CPTH method. Because of the unitarity constraint, we are forced to rewrite 
the CPTH results as an infinite geometric series using either the inverse amplitude
method or the Pad{\'e} method. The former one is more general than the latter.
 In the simplest approximation to the one loop CPTH, both method yields the
 same result. They enable us to extend the analysis of CPTH to the resonant
 region and beyond without introducing more parameters.

 On the top of the 
$\rho$ and $K^*$ masses, our calculation on the magnitudes of the form factors are
too low and only accurate to $15 \%$ in the amplitudes. 
The $\rho$ width is correctly predicted within a few percents, while that of the $K^*$
is 10 \% too high compared with the data. The r.m.s. radii are in agreement 
with the data. Extending our calculation to above 1 GeV, then the accuracy becomes
worse. The $\tau$ to $\rho$ and $K^*$ decay branching ratios are accurate to 
about 35\% but the $\tau$ to $K^+ \bar{K}^0$ and $K^+ \eta$ are too low by a factor of 2. 

In order to improve our predictions, we modify our UCPTH results which
 are obtained based on the assumption of the elastic unitarity, by incorporating the 
inelastic effect through the phenomenological polynomial ambiguity of the Omn{\`e}s
function. Mutiplying the UCPTH results on the form factors
 by the factor $(1+0.15 s/s_\rho)$ the leptonic widths of the $\rho$ and $K^*$ 
are in agreement with the data. The prediction of the branching ratios $\tau 
\rightarrow K^+ \bar{K}^0 \nu$ and $K^+ \eta \nu$
  are also in agreement with the data. The corresponding rms radii are also 
in agreement with the data.

\section{\bf Acknowledgement}

It is a pleasure for one of us (T.N.T.) to thank the hospitality of 
the Physics Department at the National High Energy Physics Laboratory K.E.K
 and in particular Prof. Y. Shimizu for hospitality. He also would like
to thank the Centre National de Recherche Scientifique for granting a 
leave of absence to visit Japan, the Hanoi National University, Hochiminh
National University
 and Hue University where part of this 
work was carried out.
(L.V.D) wants to thank the Centre de Physique Theorique de l' Ecole Polytechnique
for a fellowship and hospitality. We both thank to L. Beldjoudi for having
 participated in the intial stage of this work.

\eject

\newpage

{\bf Figure Captions}

Fig.1~: The square of the modulus of the vector pion form factor $V(s)$
as a function of the energy squared $s$. The experimental data are taken
from the review article by D. Morgan and M. Pennington
\cite{Morgan}. The solid line curve is the result of the UCPTH,
Eq.(\ref{eq:vu})~; the dashed curve is the result of the standard
one loop CPTH using the same parameter as that of UCPTH
Eq.(\ref{eq:pertv})~; the dotted curve is that obtained by simulating
the inelastic $\omega\pi$ contribution \cite{Costa}  by multiplying
the UCPTH result 
with the factor $1+0.15 s/s_{\rho}$. 

Fig.2~:The phase of the vector pion form factor in degrees as a function of the energy
squared $s$. The unitarity relation requires that it has the same
phase as that of the strong P-wave $\pi\pi$ scattering. The experimental
data for the P-wave phase shift are taken from \cite{Morgan}. The
solid curve is the calculated phase of the pion form
factor using UCPTH, Eq.(\ref{eq:vu}); the dashed curve is calculated with the CPTH,
 Eq.(\ref{eq:pertv})~; the inelastic effect due to the $\omega\pi$
contribution does no affect the phase of the UCPTH form factor below the
$\omega\pi$ threshold.

Fig.3~: Test of the convergence of the perturbation expansions. The
solid line represents the ratio $r_V$ of the one loop vector form factor
 as a function of the energy
squared $s$~; the long dashed line represents the corresponding 
ratio $r_V$ for the two loop calculation; the short dashed line
 represents the ratio $r_S$ of the scalar
form factor as a function of the energy squared $s$. The convergence
of the perturbation series requires that these ratios to be much larger
than unity.

Fig.4~: The modulus of the pion form factor squared calculated by the
UCPTH as compared with the sum of the first four terms of the Taylor
series expansion of the UCPTH calculation.

Fig.5~: The phases of the scalar pion form factor, in degrees, as a
function of the energy squared $s$.  The experimental data are taken
from the review article by D. Morgan and M. Pennington
\cite{Morgan}. The solid line curve is the result of the UCPTH,
Eq.(\ref{eq:phiu}) ; the dashed curve is the result of the standard
one loop CPTH using the same parameter as that of UCPTH,
Eq.(\ref{eq:phip}).

\newpage
\begin{figure}
\epsfbox{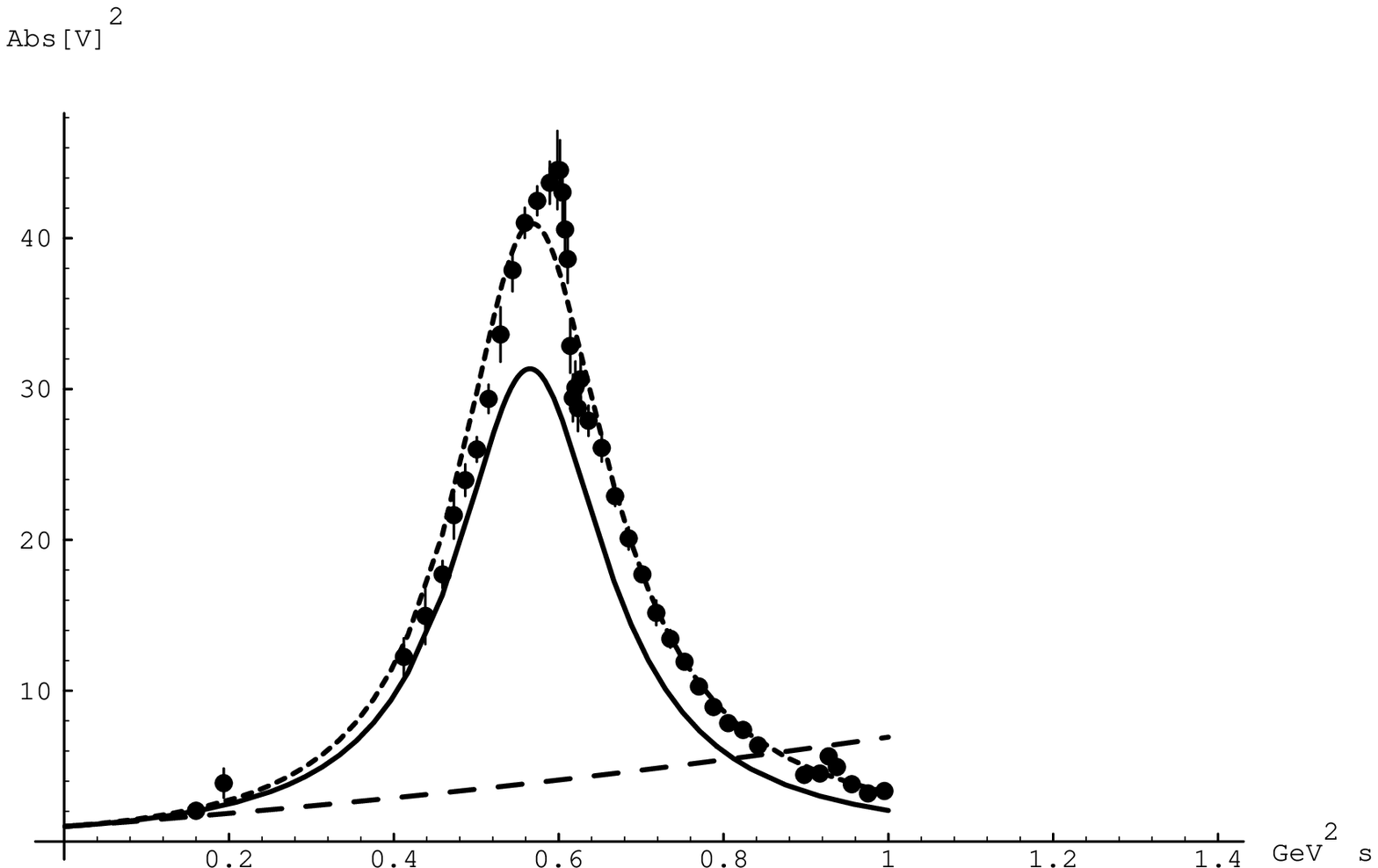}
\caption{}
\label{Fig.1}
\end{figure}
\newpage
\begin{figure}
\epsfbox{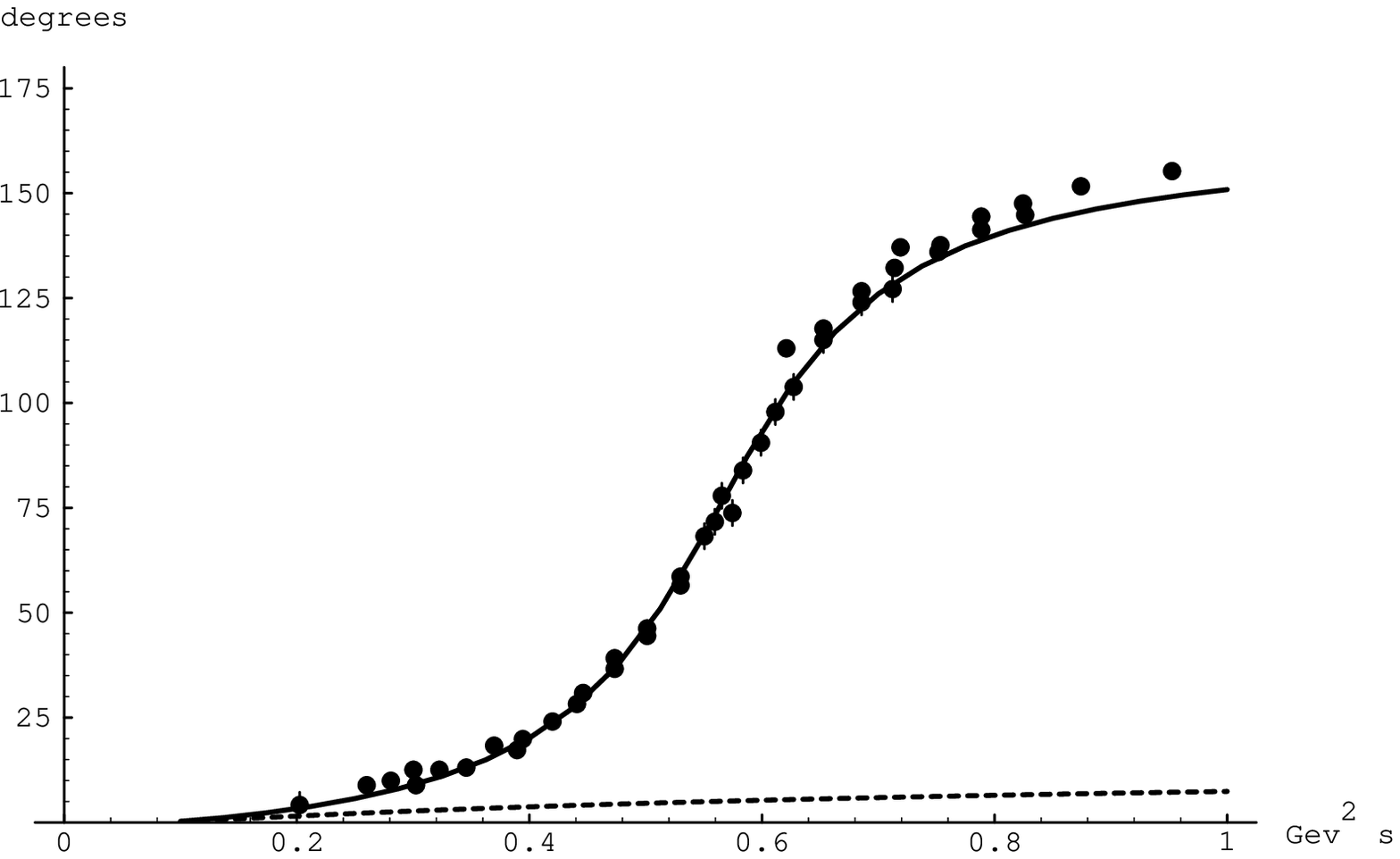}
\caption{}
\label{Fig.2}
\end{figure}
\newpage
\begin{figure}
\epsfbox{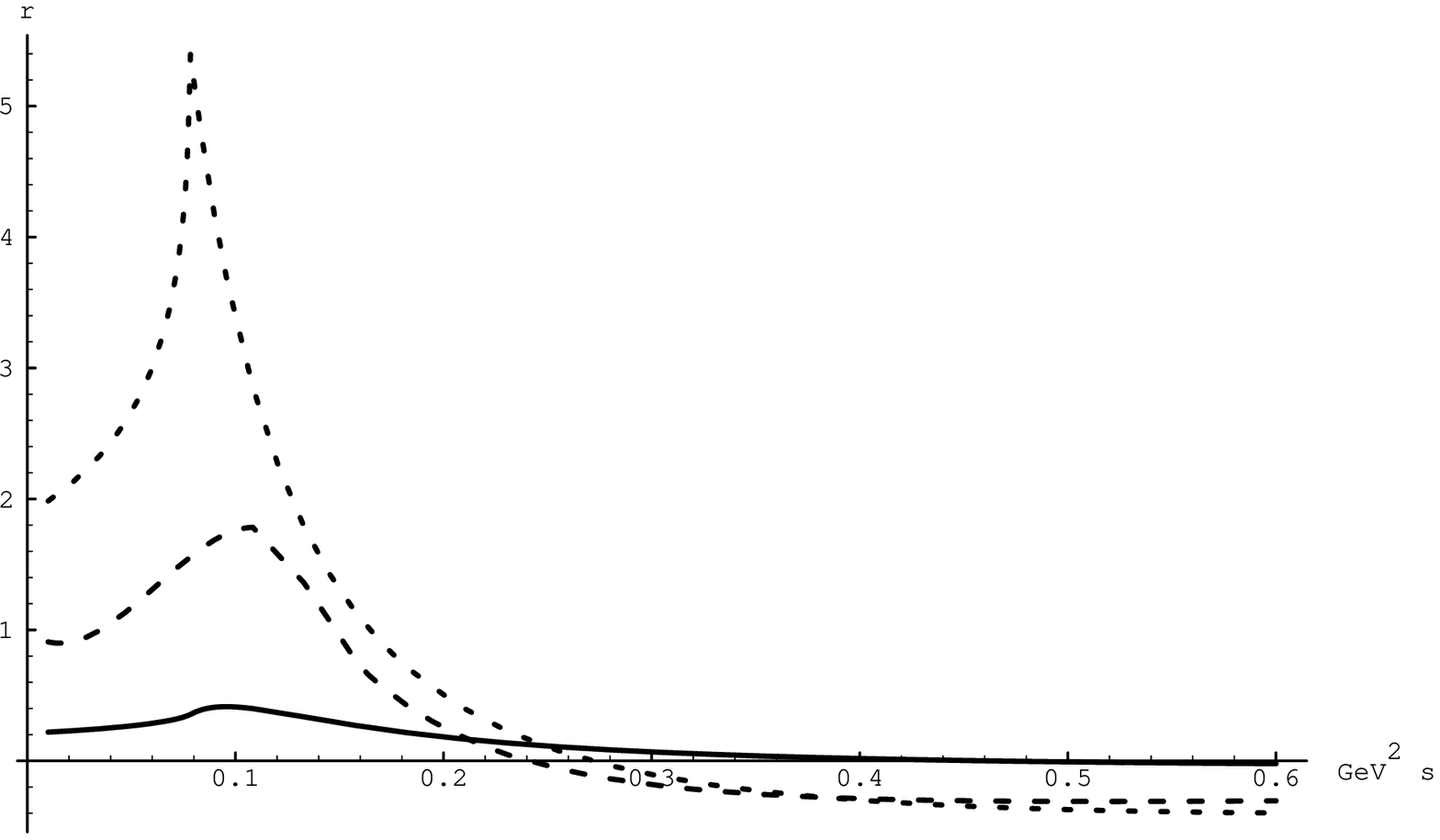}
\caption{}
\label{Fig.3}
\end{figure}
\newpage
\begin{figure}
\epsfbox{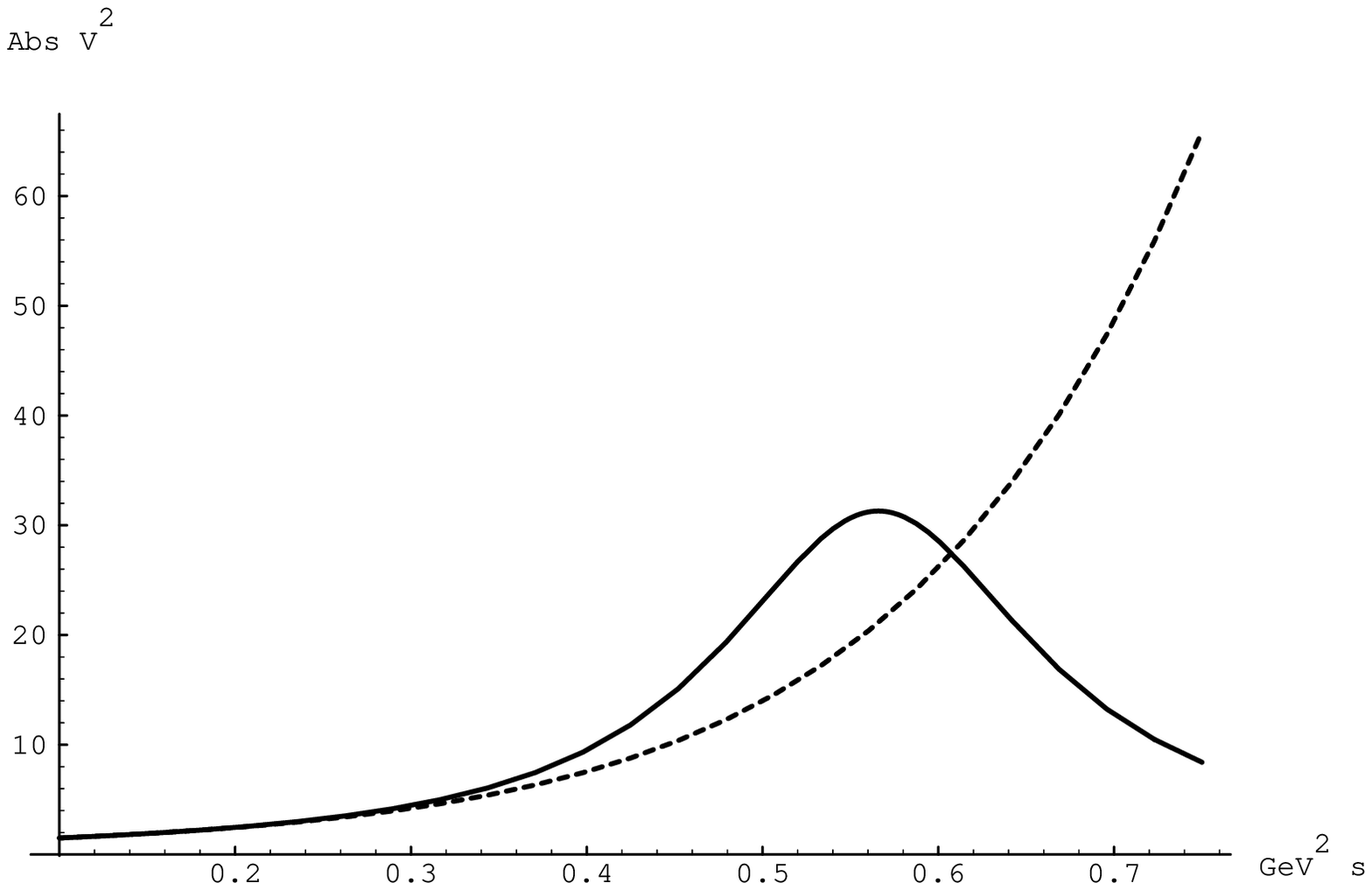}
\caption{}
\label{Fig.4}
\end{figure}
\newpage
\begin{figure}
\epsfbox{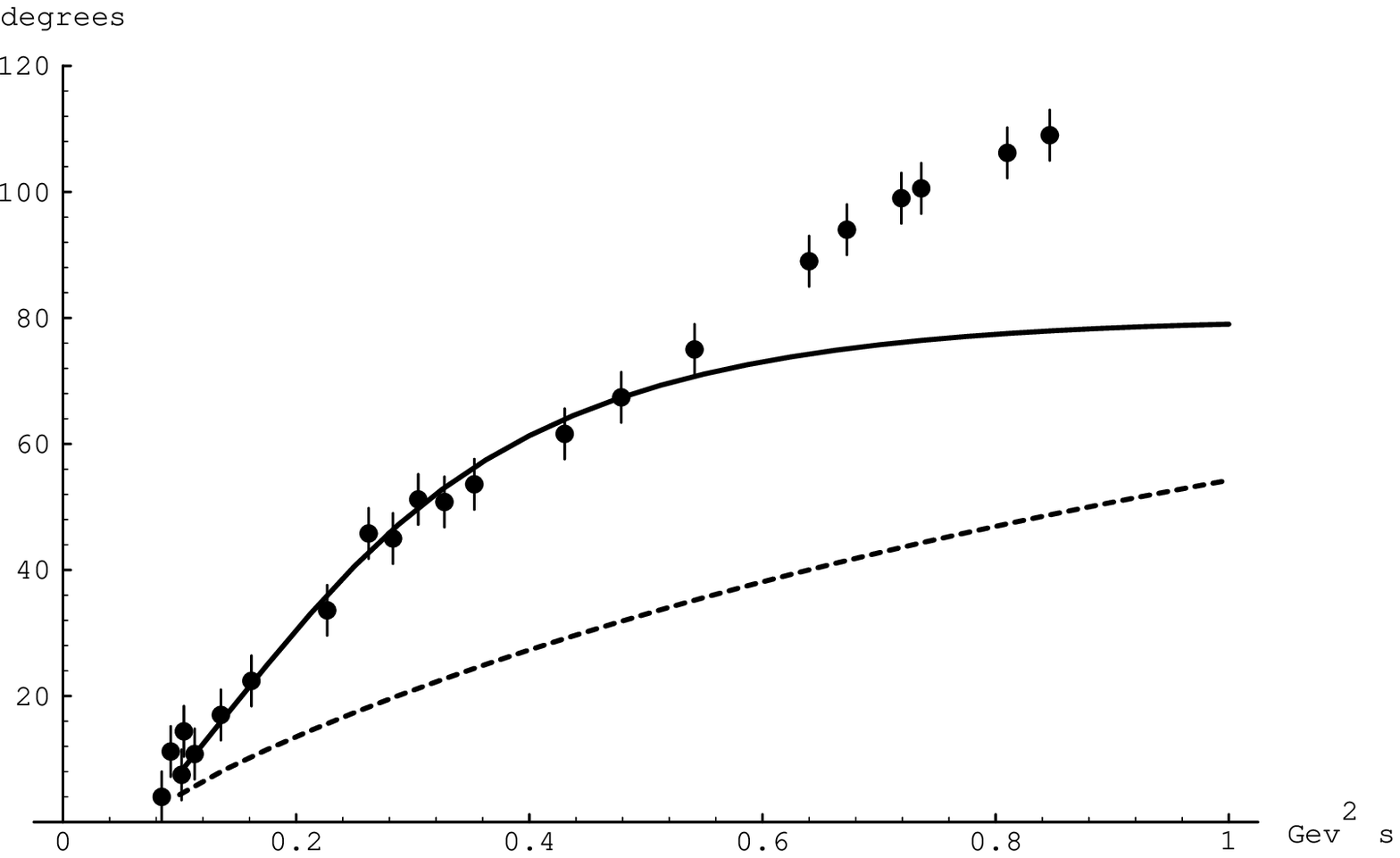}
\caption{}
\label{Fig.5}
\end{figure}

\end{document}